\documentclass[%
 twocolumn,
superscriptaddress,
 amsmath,amssymb,
 aps,
prb,
]{revtex4-1}
\usepackage{graphicx,xcolor}
\usepackage{mathtools}
\usepackage{hyperref}
\usepackage{comment}
\usepackage[utf8]{inputenc}
\usepackage{feynmp}
\usepackage{pgfplots}
\def\Gbar{{\bar{G}}}

\def\hbar{{\bar{h}}}

\def\Gscript{{\cal G}}
\def\Gbar{{\bar G}}
\def\beq{\begin{equation}}
\def\eeq{\end{equation}}

\def\beqa{\begin{eqnarray}}
\def\eeqa{\end{eqnarray}}

\begin{document}
\title{%LSMS Embedding \\ or \\
%LSMS embedding scheme for disordered electron systems
Locally self-consistent embedding approach
for disordered electronic systems}
%\textcolor{red}{here we are not doing what people refer to the Multiple scattering theory}}
\author{Yi Zhang}
\email{zhangyiphys@gmail.com}
\affiliation{Department of Physics \& Astronomy, Louisiana State University, Baton Rouge, LA 70803, USA}
\affiliation{Center for Computation \& Technology, Louisiana State University, Baton Rouge, LA 70803, USA}
\author{Hanna Terletska}
\affiliation{Department of Physics and Astronomy, Middle Tennessee State University, Murfreesboro, TN 37132, USA}
\author{Ka-Ming Tam}
\affiliation{Department of Physics \& Astronomy, Louisiana State University, Baton Rouge, LA 70803, USA}
\affiliation{Center for Computation \& Technology, Louisiana State University, Baton Rouge, LA 70803, USA}
\author{Yang Wang}
\affiliation{Pittsburgh Supercomputing Center, Carnegie Mellon University, PA 15213, USA}
\author{Markus Eisenbach}
\affiliation{Center for Computational Sciences, Oak Ridge National Laboratory, Oak Ridge, TN 37831, USA}
\author{Liviu Chioncel}
\affiliation{Theoretical Physics III, Center for Electronic Correlations and Magnetism, Institute of Physics, University of Augsburg, D-86135 Augsburg, Germany}
\affiliation{Augsburg Center for Innovative Technologies, University of Augsburg, D-86135 Augsburg, Germany}
\author{Mark Jarrell}
\affiliation{Department of Physics \& Astronomy, Louisiana State University, Baton Rouge, LA 70803, USA}
\affiliation{Center for Computation \& Technology, Louisiana State University, Baton Rouge, LA 70803, USA}

\date{\today}

\begin{abstract}
We present a new embedding scheme for the locally self-consistent method to study disordered electron systems. 
We test this method in a tight-binding basis and apply it to the single band Anderson model. %
%We demonstrate that embedding significantly improves system size convergence and hence reduces the computational costs as expected.
%
The local interaction zone is used to
efficiently compute the local Green's function
of a supercell embeded into a local typical medium.
We find a quick convergence as the size of the local interaction zone which reduces the computational costs as expected.
This method captures the Anderson localization transition and accurately predicts the critical disorder strength.
The present work opens the path towards the development of a typical medium embedding scheme for the $O(N)$ multiple scattering methods.  
\end{abstract}

\maketitle

%***************************************************
\section{Introduction}
%*************************************************
%{\bf{Disorder and why it is important}}
Disorder which is ubiquitous feature of real materials (in the form of impurities or defects in perfect crystals, or chemical substitutions in alloys and random arrangements of electron spins or glassy systems) plays a key role in changing and controlling their properties~\cite{e_abrahams_10,r_elliott_74,d_belitz_94,d_vollhardt_84,tmdca_review}. %It has been found to cause dramatic changes to atomic, magnetic and phonon phenomenon systems. Especially, a strong disorder, which can lead to electron localization, known as Anderson localization~\cite{p_anderson_1958}. 
As shown long ago by Anderson~\cite{p_anderson_1958}, disorder in atomic coordinates creates spatially confined or ``localized'' electron eigenstates near the Fermi level. 
Electron localization has been found to play a crucial role in a number of materials, starting from the prototype
%This includes electron localization in 
two-dimensional electron systems~\cite{mosfet_2010}, displaying metal to insulators transitions~\cite{e_abrahams_10}, to various well known semiconducting materials including
% the nowadays fashionable 
Dirac~\cite{Neupane2014,Liu864,Borisenko_14,Liu2014} and Weyl~\cite{Lv_2015,Xu613,Xu2015} semi-metals. 

Among the well known studied system, are semiconductors  such as Si doped with P, B, S or Ti. For instance in Si:P
the P donors sit substitutionally on the Si sites and 
for low concentrations, according to Mott~\cite{mott.67,Mott_1968}, there is a 
negligible overlap between the wave functions of the 
donor electrons, and the material is an insulator. At high
concentrations when the overlap is large compared with the
on-site repulsion the material is a metal. These observations
lead Mott~\cite{mott.67,Mott_1968} to formulate a
phenomenological theory for the transition from the
insulating to the metallic state (localized to itinerant electrons) in terms of a critical concentration $n_c$ and the
average distance between the impurities fulfilling the 
relation: $n_c^{1/3} a_B \approx 1/4$, where $a_B$ is the spatial extension (effective Bohr radius) of the P donor
electrons. 
The alternative view due to Anderson~\cite{p_anderson_1958}, 
involves localization due to random one-electron potentials 
seen by the electrons. For low donor concentrations the 
one-electron energy spread in the random potentials 
(energy distance between consecutive energy eigenstates) is large compared 
with the energy band-width and the electronic states are 
localized. At high concentrations, the localized impurity states form the impurity band and the extended states appear
separated from the localized states by a mobility edge. The metal-insulator transition can also happen by doping which shifts the Fermi level across the mobility edge.

Dilute magnetic semiconductors (with a subtle interplay between magnetism and electron localization) and intermediate band photovoltaics (which hold the promise to significantly improve solar cell efficiency) are among another important class of materials where disorder plays a fundamental role on their properties. 
Dramatic improvement in crystal growth in recent years has enabled
preparation of samples with a significant control over the degree of disorder. 
For example, localization has been definitively seen in single crystals
of Li$_x$Fe$_7$Se$_8$ 
%as a consequence of disorder associated with Li doping.  
despite of a finite density of states at the Fermi energy 
confirmed via specific heat and reflectivity measurements~\cite{Ying2016}. 
Besides these systems, thermoelectric and topologic materials such as
Cd$_{3}$As$_{2}$, Na$_{3}$Bi~\cite{Neupane2014,Liu864,Borisenko_14,Liu2014} and TaAs, NbAs~\cite{Lv_2015,Xu613,Xu2015} hint towards the presence of significant disorder effects that still remain to be fully understood.

In recent decades, the {\em ab initio} methods based on density functional theory (DFT)~\cite{Hohenberg1964,Kohn1965,kohn.99,jo.gu.89,jone.15} 
have become the most important tool to calculate properties of ordered 
crystalline solids. The band-theory as such can not be used to treat disordered
solids because of the lack of translation invariance. For modeling disordered
solids calculations were performed using effective medium theories, among them 
the Coherent Potential Approximation (CPA)~\cite{p_soven_67} proved to be a simple 
and transparent theory that is able to capture important features of the electronic
structure of alloys. The CPA has been used to solve models which helped in 
providing physical interpretation of experimental results on real alloys. 
To become quantitative the CPA equations have been formulated for the muffin-tin
potentials within the multiple-scattering Korringa-Kohn-Rostocker (KKR)
method~\cite{Korringa1947,Kohn1954}. The configurational average could be 
performed 
over the scattering path operator, instead of the Green's function (used
for models), simplifying the implementation of the CPA for materials 
calculations~\cite{st.wi.71}.
Later the CPA was also implemented within the linearized muffin-tin
orbitals~\cite{ande.75} (LMTO) basis
set~\cite{ku.dr.87,ku.dr.90,ab.ve.91,ab.sk.93}. With the advent of the third-generation exact 
muffin-tin orbitals~\cite{an.je.94.2,vito.01,vitos.10} (EMTO) method, and the 
full-charge density~\cite{vi.ko.94} (FCD) technique, it was possible to go beyond
the atomic-sphere approximation (ASA) with CPA calculations~\cite{vi.ab.01}, and
investigate the energetics of anisotropic lattice distortions.

\begin{figure}[htb]
\includegraphics[width=0.5\textwidth]{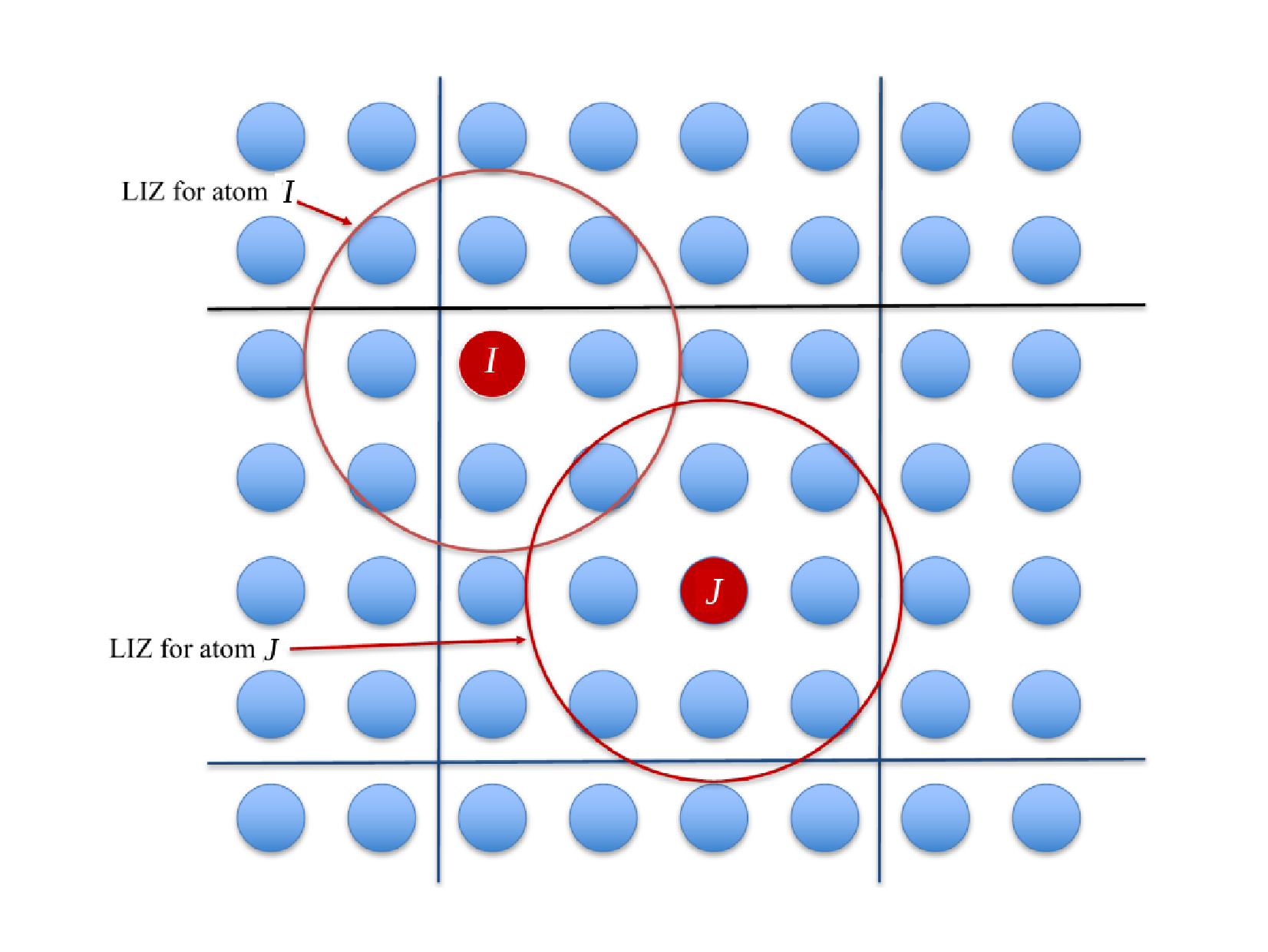}
\caption[]{A schematic representation of LIZ (red circles), 
centered around the sites $I$, $J$.}
\label{fig:liz}
\end{figure}

An alternative to the effective medium theories are supercells calculations 
which nowadays can be performed on systems containing thousands
of atoms. This has been made possible by the development of order-N (O($N$)) 
methods based on plane wave expansions~\cite{ca.pa.85,pa.te.92} or multiple
scattering theory~\cite{Wang1995}.  
In the multiple scattering formulation the linear scaling {\em ab initio} method
came to be known as the Locally Self-consistent Multiple Scattering (LSMS) method.
LSMS achieves linearly scaling for very large systems with up to tens of thousands of atoms, via the introduction of a smaller local interaction zone (LIZ) of size $N_{\rm LIZ}$ of several hundreds of sites. Within the LIZ the electronic structure problem is solved explicitly with free space boundary conditions. In LSMS analysis, as the LIZ moves through each site, the explicit DFT solution introduces correlations due to the different disorder configuration (Fig.~\ref{fig:liz}). 

In this manuscript, we present our concept of embedding that combines
the real space construction with the momentum space self-consistency. 
In this scheme we follow the ideas of LSMS implementation~\cite{Wang1995} and investigate the electrons 
localization in a three-dimensional (3D) Anderson model. 
This represents a natural extension of our previous study on multiple
scattering formulation to the problem of Anderson localization~\cite{tmdca_review}. However, 
contrary to standard LSMS, our new scheme does not explicitly rely on the multiple scattering aspect of the cluster solver.
In comparing the results obtained using the local average Coherent Potential and Typical effective medium embedding schemes, we find that the typical medium embedding plays a significant role in capturing electron localization.\cite{ab.ni.96,ab.si.97} 
In particular, the extrapolated critical disorder strength for 
Anderson transition is in excellent agreement with the known 
literature results~\cite{Slevin99,Bulka85,Fehske,song_07,slevin_01,romer_10,romer_11,Mackinnon_81}. 

The remainder of this paper is organized as follows. After the
introduction, we discuss some conceptual details of LSMS in 
Sec.~\ref{sec:LSMS}, followed by the typical medium formulation of 
the Anderson localization, Sec.~\ref{sec:TM}. 
In Sec.~\ref{sec:LSMS_intro}, we present the computational details
and the self-consistent loop used in the present calculations. 
Then in Sec.~\ref{sec:results} we illustrate the effective medium embedding using the Hamiltonian formulations and present results  
for the coherent potential effective medium and the typical medium.

\section{Calculating properties with LSMS}
\label{sec:LSMS}

LSMS has the unique capability for studying extremely large and 
disordered systems.\cite{Eisenbach2017,Yang2017} Yet, the
standard construction of the LIZ using an open boundary condition
limits the applicability of LSMS to the description of disordered
metals only, and in particular fails to properly describe band gaps 
and electron localization. 
This is because free space boundary conditions, for which the potential
is set to be zero, couple the LIZ to a free space density of states 
which increases as a square root for positive energies, so that the gaps in a semiconducting system are filled in and blurred, and semiconductors appear to be metals.
One possible way to overcome this issue is to use self-consistently 
determined boundary conditions. 
Inspired by the LSMS construction, Abrikosov {\it et. al.}~\cite{ab.ni.96,ab.si.97}
suggested a locally self-consistent method in which the LIZ size is reduced
by considering an effective scatterer outside the LIZ. The choice made for this
scattering matrix is the CPA single-site $t$-matrix. The excellent convergence achieved through this method~\cite{ab.ni.96,ab.si.97,ab.ve.91} allowed also to 
address the problems of total energy calculations in alloys~\cite{jo.pi.93,pins.98,uj.fa.00}.

Note that the open boundary conditions imposed upon LIZ also disable
the ability of LSMS to study electron localization. 
It has been shown recently~\cite{v_dobrosavljevic_03,tmdca_review} 
that the average density of states is not critical through the Anderson transition, and instead the typical (geometrically averaged density of
states) needs to be used to identify the transition~\cite{v_dobrosavljevic_03}. Therefore, in order to capture the electron localization in real materials using the 
LSMS scheme, the boundary conditions must couple the LIZ to an 
effective medium which reflects the typical order parameter. 
The essential aspects of the typical medium theory are discussed in 
the following section.

\section{Typical medium approach}
\label{sec:TM}
Recent studies~\cite{v_dobrosavljevic_03,tmdca_review} have shown 
that disorder-driven electron localization is captured by typical medium approaches. 
The typical medium approach takes into account the dramatic changes of 
the distribution of the local density of states (LDOS)
through the localization
transition. More specifically, it changes from a Gaussian distribution 
to a skewed log-normal distribution, where the algebraic average of the LDOS stays finite while the geometric average of the LDOS which is usually called the typical density of states (TDOS) drops to zero~\cite{g_schubert_10}. This property of TDOS makes it a potential candidate order parameter for the localization transition. 
Typical medium analysis helps to overcome the shortcomings of the standard effective medium methods such as CPA~\cite{p_soven_67} and DCA~\cite{jarrell_2001} which fail to describe the localization transition.  
The typical medium theory (TMT) introduced for the first time by 
{\it Dobrosavljevi\`{c} et. al.}~\cite{v_dobrosavljevic_03},  successfully captures precursors of the Anderson localization transition, but strongly overestimates the localization effect 
due to its single site nature. Later, a finite size cluster extension of TMT, the so called TMDCA, was introduced~\cite{tmdca_review} which 
accurately predicts the critical disorder strength of the Anderson localization transition in a single band Anderson model with uniform disorder. 
The TMDCA has been extended to systems with off-diagonal disorder~\cite{h_ter_14} and to multi-band systems~\cite{y_zhang_15} 
in model Hamiltonians, where it accurately reproduces the localization phase diagrams. The obtained results are in agreement with other well established theoretical techniques such as the transfer matrix and 
the kernel polynomial methods~\cite{h_ter_14,y_zhang_15}. More recently, TMDCA was also combined with the first-principle calculations
to study the localization effects in realistic materials with disorder~\cite{y_zhang_17,ekuma_17,y_zhang_18}. The TMDCA formulated within the multi-scattering theory is still used at the 
model Hamiltonian level~\cite{h_ter_17}.

\section{Embedding scheme}
\label{sec:LSMS_intro}

In this section, we describe the construction of the effective medium
embedding scheme using concepts of LSMS. The self consistent 
embedding is a coupling framework which provides rigorous boundary
conditions for the primary region  (site, or cluster) to be embedded
into a larger self-consistently determined environment. 
Central to the embedding theory is the embedding potential (e.g. a
self-energy) which embodies the functional connection between the primary region and the environment.
In the original LSMS calculation no embedding scheme is used, in 
other words, the LIZ was effectively embedded in a vacuum~\cite{Wang1995}. Latter it has been shown that the size of the
LIZ and hence the computational effort may be considerably 
reduced by embedding the LIZ in an effective medium~\cite{ab.ni.96}. 
Using the CPA as the embedding effective medium leads to the so-called 
locally-self-consistent Green’s function method~\cite{ab.si.97} which
was applied to the metallic alloys. 
On contrary, as will be shown below, the embedding into the effective typical medium allows to address the Anderson localization transition.

\begin{figure}[htb]
\includegraphics[width=0.49\textwidth,viewport=15 120 770 480,clip]{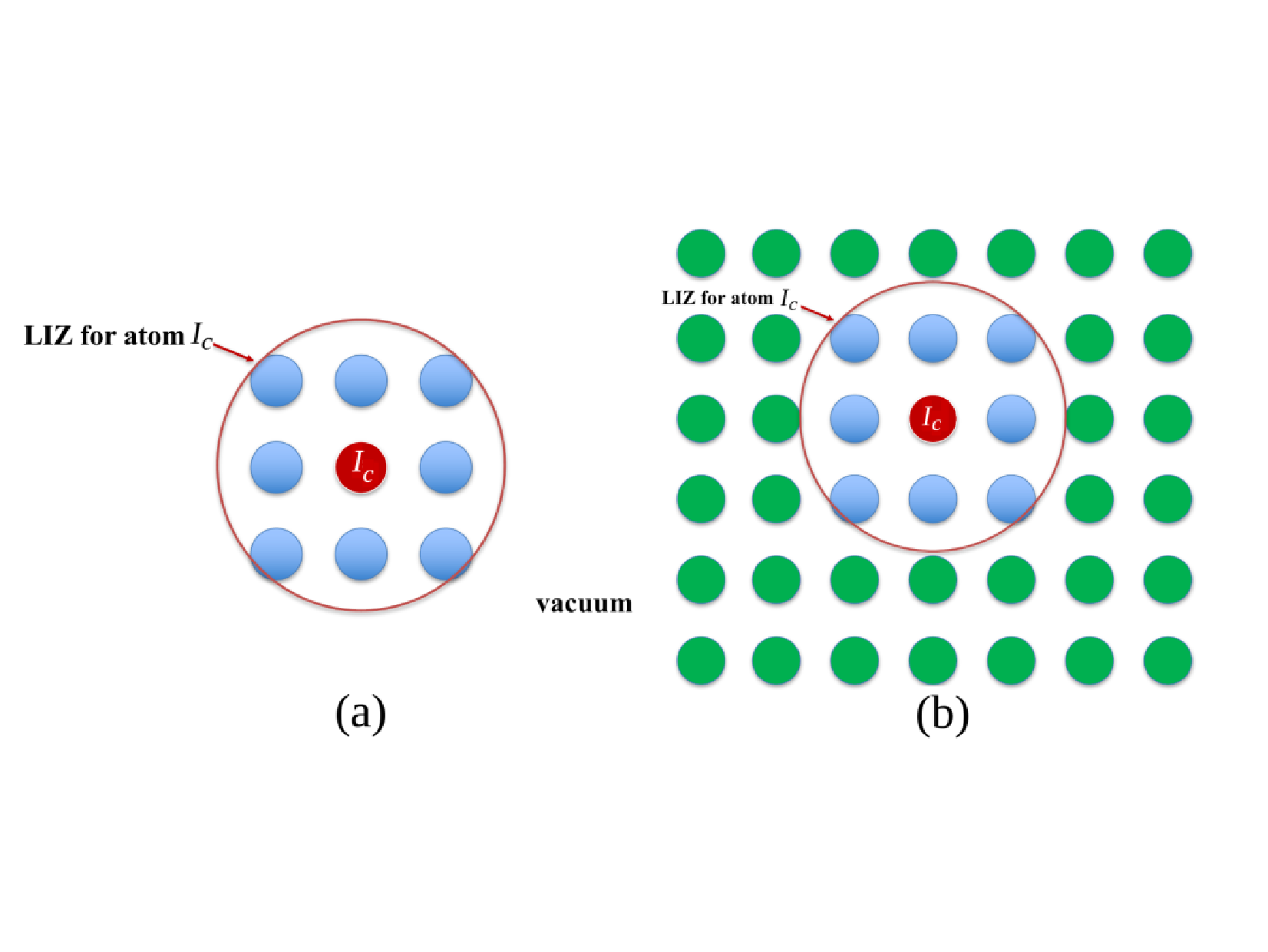}
\caption[]{Setups of LIZ for conventional LSMS without embedding (a), and with embedding scheme (b). The index $I_c$ denotes the center of the LIZ. Blue circles represent sites inside the LIZ while the green circles represent the homogeneous effective medium.}
\label{fig:lsms_embedding}
\end{figure}

In the following we describe our computational scheme.
We first surround a site $I_c$ of the lattice and form
the LIZ (red circle of Fig.~\ref{fig:lsms_embedding}). 
Sites within the LIZ are denoted by capital letters $(I,J)$. 
We choose the LIZ of linear dimension $L_{\rm LIZ}$ to be 
contained in a supercell of dimension $L_c > L_{\rm LIZ}$.
\footnote{Note however that within LSMS no such requirement is 
necessary, even a single-site can be considered as supercells for a
specific large LIZ.}
The local interaction zone will be moved through all sites of the 
supercell. 
The supercell is repeated to restore the lattice translation
invariance, generating the set of $K$-points. Given a supercell 
Green's function $\Gbar(\omega,K)$ the Fourier transform provides 
the real space Green's function in the supercell:
\begin{equation}~\label{eq:Gliz_m}
\Gbar_{IJ}(\omega) = \frac{1}{N_c}\sum_K e^{iK\cdot R_{IJ}} \Gbar(\omega,K)
\end{equation}
where $K$ are the supercell wavenumbers, with $N_c$ the number of 
sites in the supercell, defined in the same way as those in 
DCA~\cite{jarrell_2001}. The indices $(I,J)$ cover all lattice sites within the supercell.
The corresponding LIZ Green's function has the form:
\begin{equation}~\label{eq:Gliz_m}
\Gbar_{IJ}^{\rm LIZ}(\omega) = \frac{1}{N_c}\sum_{K} e^{iK\cdot R_{IJ}} \Gbar(\omega,K), \ \  (I,J)\in {\rm LIZ}
\end{equation}
 where $\Gbar(\omega,K)$ is  
defined through the coarse-graining procedure as:
\begin{equation}~\label{eq:Gliz_m0}
\Gbar(\omega,K)=\frac{N_c}{N}\sum_{\tilde{k}} \frac{1}{\omega - \epsilon(K+\tilde{k}) -\Sigma_l(\omega)}.
\end{equation}
The local effective self-energy is denoted by $\Sigma_l(\omega)$  and  $\epsilon(K+\tilde{k})$ is the lattice dispersion.
The supercell wavenumbers $K$ correspond to the $N_c$ cells that divide
the first Brillouin zone equally. The wavenumbers $\tilde{k}$ label 
the wavenumbers within each cell surrounding $K$.
The supercell is embedded into the effective medium, represented by
$\Sigma_l$ present at all supercell sites. Therefore, LIZ sites
(contained in the supercell) experience the presence  
of the effective medium. Consequently, we may rewrite the 
real space LIZ Green's function $\Gbar^{LIZ}(\omega)$ in the 
following form: 
\begin{equation}
\underline{\Gbar}^{\rm LIZ}(\omega) = \left(\omega\cdot\underline{\mathbb{I}} - \underline{t'} - \Sigma_l(\omega)\cdot\underline{\mathbb{I}} - {\underline{\Gamma}^{\rm LIZ}(\omega)} \right)^{-1},
\end{equation}
The underline indicates matrices of dimension $L_{\rm LIZ} \times L_{\rm LIZ}$ 
corresponding to the number of sites contained within the LIZ. The hopping 
matrix elements within the LIZ are given by ($\underline{t'}$), and $\underline{\mathbb{I}}$ is the corresponding identity matrix. 
The hybridization function between the LIZ and the effective medium is
$\underline{\Gamma}^{\rm LIZ}(\omega)$ and the LIZ excluded Green's function can 
be defined as: 
\begin{equation}
\underline{\Gscript}(\omega) =\left(\omega\cdot\underline{\mathbb{I}} - \underline{t'} - {\underline{\Gamma}^{\rm LIZ}(\omega)} \right)^{-1}
\label{eq:gscrip}
\end{equation}
We do not explicitly evaluate the hybridization function
${\underline{\Gamma}^{\rm LIZ}(\omega)}$, instead we directly calculate the LIZ 
excluded Green's function using: 
\begin{equation}\label{g_tilde}
\underline{\Gscript}^{-1}(\omega) =\left(\underline{\Gbar}^{\rm LIZ}(\omega)\right)^{-1}+\Sigma_l(\omega)\cdot\underline{\mathbb{I}}
\end{equation}
Here $\underline{\Gscript}(\omega)$ is the real space Green's function 
inside LIZ in the absence of disorder.  
Accordingly,  
$\underline{\Gscript}(\omega)$ takes the same values for all possible
LIZs obtained by running the centra of LIZ ($I_c$) through each sites
in the supercell. 
Within the supercell we include the disorder potential, and we
calculate the  Green's function within each LIZ  
centered around the site $I_c$:
\begin{equation}
\left(\underline{G}^{\rm LIZ}(\omega,V,I_c)\right)^{-1} = \underline{\Gscript}^{-1}(\omega) - \underline{V}(I_c)
\end{equation}
where $\underline{V}(I_c)$ is a diagonal matrix of size $N_{LIZ}$.
Note, that the matrix $\underline{G}^{\rm LIZ}(\omega,V,I_c)$  
has the same dimension. The index $I_c$ (the center of the LIZ 
seen in Fig.~\ref{fig:lsms_embedding}) serves also as an additional
label indicating the presence of disorder at that specific site.

We average $\underline{G}^{\rm LIZ}(\omega,V,I_c)$ over the different
LIZs realizations within the supercell, which is expressed as
$\frac{1}{N_c}\sum_{I_c}(\cdots)$ and over the disorder configurations which 
is indicated by the angle brackets $\langle \cdots \rangle_V$.
We employ two types of averaging, for the effective medium:
\begin{itemize}
    \item Linear average (CPA)
\begin{equation}
\underline{G}_{ave}^{\rm LIZ}(\omega) = \frac{1}{N_c}\sum_{I_c}\left\langle \underline{G}^{\rm LIZ}(\omega,V,I_c) \right\rangle_V\,.
\label{eq:averageGLIZ}
\end{equation} 
and for the typical medium:
\item Typical average (TMT)
\begin{equation}\label{eq:ansatz}
\begin{split}
\underline{G}_{typ}^{\rm LIZ}(\omega)=&e^{\frac{1}{N_c}\sum_{I_c}\left\langle \ln\left(\rho_{I_c I_c}(\omega,V,I_c)\right)\right\rangle_V }\\
&\times\frac{1}{N_c}\sum_{I_c}\left\langle \frac{\underline{G}^{\rm LIZ}(\omega,V,I_c)}{{\displaystyle {\textstyle {\scriptstyle \rho_{I_c I_c}(\omega,V,I_c)}}}}\right\rangle_V
\end{split}
\end{equation}
where the density $\rho_{I_cI_c}(\omega,V,I_c)$ is the density at the center of the LIZ defined as
\begin{equation}
    \rho_{I_c I_c}(\omega,V,I_c)=-\frac{1}{\pi}{\rm Im}(\underline{G}^{\rm LIZ}(\omega,V,I_c))_{I_c,I_c}
\end{equation}
To obtain the typical value, we perform a geometric average of the local density of states at the center of the LIZ over all the LIZs and disorder configurations, which is expressed as the exponential term in Eq.~\ref{eq:ansatz},  
while the second multiplicative term of Eq.~\ref{eq:ansatz} is a linear average of the whole LIZ Green's function that are normalized by the density of states at its center.
\end{itemize}

\begin{figure}[htb]
\includegraphics[width=0.5\textwidth]{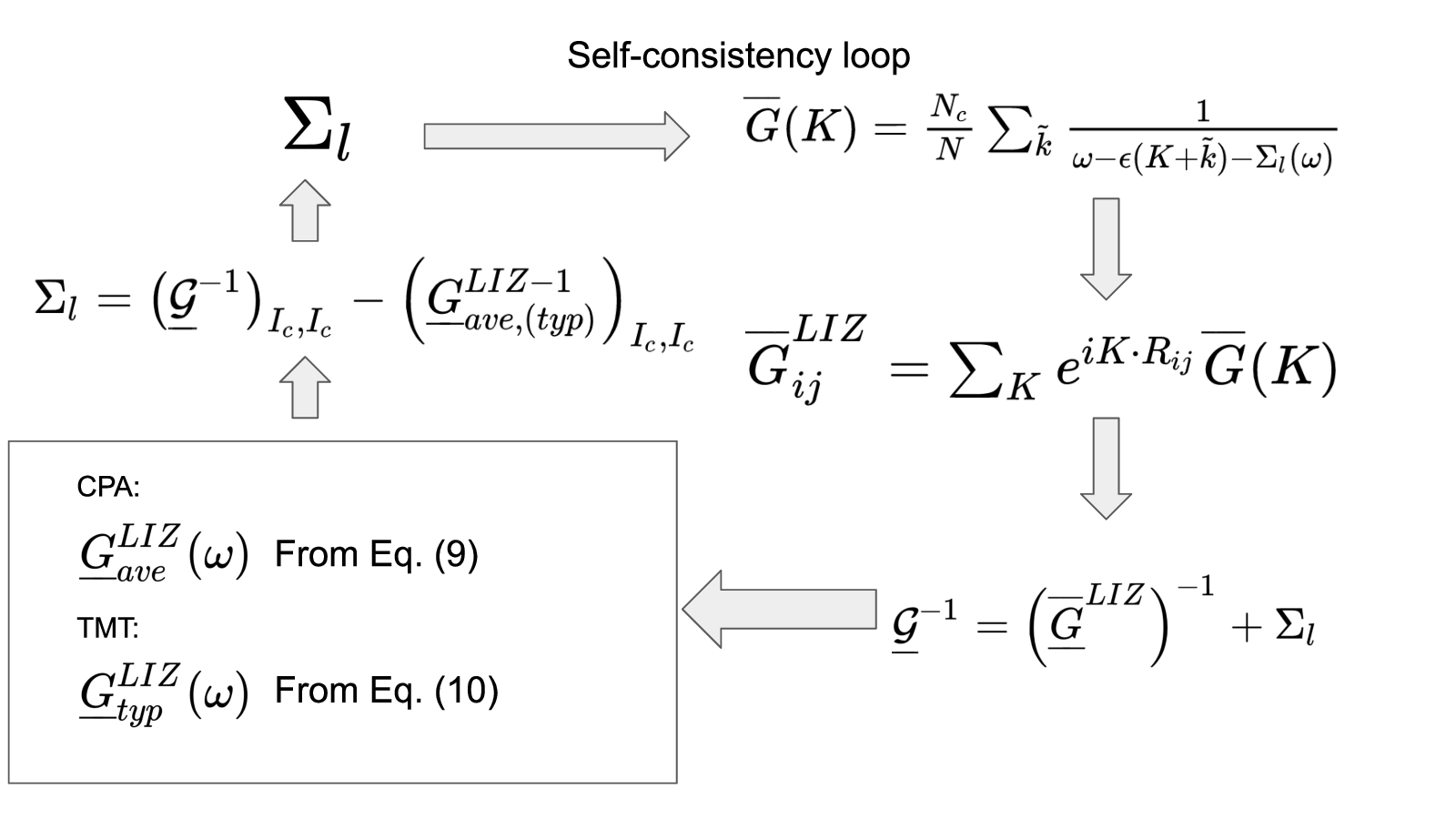}
\caption[]{The self-consistency loop for CPA/TMT embedding method.}
\label{fig:self_loop}
\end{figure}

In Fig.~\ref{fig:self_loop} we present the self-consistency loop for
both CPA and TMT embeddings as described above. The central quantity 
to be iterated within the self-consistent calculation is the local
effective self-energy $\Sigma_l(\omega)$. This can be computed as:
\begin{equation}
\Sigma_l(\omega) = \left({\underline{\Gscript}}^{-1}(\omega)\right)_{I_c,I_c}
- \left({\underline{G}_{ave(typ)}^{\rm LIZ-1}}(\omega)\right)_{I_c,I_c}
\label{eq:Sigmalocal}
\end{equation}
where $\underline{G}_{ave(typ)}^{\rm LIZ}(\omega)$ is the disorder averaged $\underline{G}^{\rm LIZ}(\omega,V)$.
When constructing the typical Green's function for the TMT embedding,
we replace $\underline{G}_{ave}^{\rm LIZ}(\omega)$ by $\underline{G}_{typ}^{\rm LIZ}(\omega)$ in the calculation of the 
local self energy $\Sigma_l(\omega)$ (Eq.~\ref{eq:Sigmalocal}). 
In Eq.~\ref{eq:ansatz}, the sum is over the sites $I_c$ in the supercell. 
The typical Green's function is constructed in the spirit that its imaginary part
gives the geometric average of the LDOS for the central sites in all the LIZs in all the
disorder configurations. This quantity called the typical density of states (TDOS)
serves as an order parameter to describe the localization transition~\cite{v_dobrosavljevic_03}. 
In the limit of weak disorder the difference between TDOS and normal DOS is 
negligible and Eq.~\ref{eq:ansatz} reduces to the normal averaged Green's function
Eq.~\ref{eq:averageGLIZ}, consequently the TMT embedding reduces to the CPA embedding.

Note that a full matrix inversion is required only within the LIZ.  Thus, both the CPA and the TMT-based LIZ algorithms scale like $N_c N_{\rm LIZ}^3$, where the prefactor is due to the need to solve Eq.~\ref{eq:Sigmalocal} at the LIZ centered on every site in the system. 

\section{Results}
\label{sec:results}

In the following we apply the above embedding method to the 
single band 3D Anderson model with the Hamiltonian:
\begin{equation}
H=-t\sum_{<ij>\sigma}(c_{i\sigma}^{\dagger}c_{j\sigma}+h.c.)+\sum_{i\sigma}V_{i}n_{i\sigma}\,.
\label{eq:H}
\end{equation}
The first term describes electrons with spin $\sigma$, hopping
with the amplitude $t$, between sites $i$ and $j$ (only
nearest-neighbor hopping is included). The second term describes
static scattering processes on the local disorder center. The local potential $V_{i}$ is modeled as 
the random number drawn form a uniform box distribution, $p(V)={\displaystyle \frac{1}{2W}\Theta(W-|V|)}$. By the condition $4t=1$ the energy units are fixed. 
In the calculation, we choose cubic supercells of size $N_c=L_c^3$. The
corresponding LIZ volumes are also of cubic shape with size $N_{\rm LIZ}=L_{\rm LIZ}^3$.
The calculations were performed for a total of 400 disorder realizations with the proper disorder averaging. 
Calculations are performed in the thermodynamic limit despite 
the finite size of the LIZ. This is a consequence of the fact 
that the supercell containing LIZ is embedded into an effective
medium. Note that embedding schemes leads to a faster convergence
rates in size for both supercell as well as for LIZ~\cite{ab.ni.96}. 

\begin{figure}[h]
\includegraphics[width=0.45\textwidth]{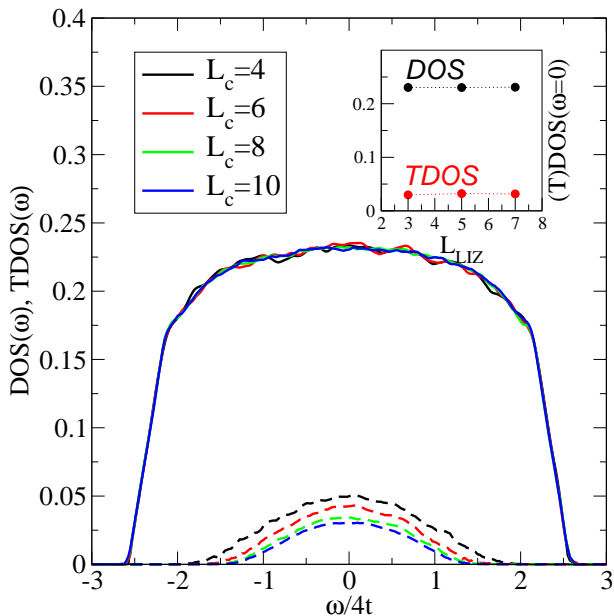}
\caption[]{Comparison of the average DOS (solid curves) and the typical TDOS (dashed curves) at different supercell sizes L$_c$= 4,6,8,10 with fixed LIZ size L$_{\rm LIZ}$=3. The disorder strength is set to be $W=2.0$.
Inset: Comparison of DOS and TDOS at the band center with supercell size L$_c$=10 and three different LIZ sizes: 3, 5, 7. They are independent of the size of LIZ. }
\label{fig:dos_tdos_liz}
\end{figure}

Fig.~\ref{fig:dos_tdos_liz} shows the comparison of average and
typical DOS for different supercell sizes. We also plot in the
inset the DOS and TDOS at the band center as a function of LIZ 
size. As the size increases from 3 to 7 both DOS and TDOS remain 
almost unchanged, therefore in the following we will show only
results obtained for L$_{\rm LIZ}$=3.

The other relevant length scale is the size of the supercell
L$_c$. We performed calculations for different supercell sizes 
keeping fixed L$_{\rm LIZ}$=3, as shown in the main panel of 
Fig.~\ref{fig:dos_tdos_liz}.
As can be seen DOS results shows no significant change for 
different supercell sizes. On contrary the magnitude of TDOS 
decreases slowly with the increase of the supercell size L$_c$.
Since the TDOS defines the order parameter of the Anderson 
localization, it vanishes at the critical transition 
point~\cite{tmdca_review}. 
In order to compute the value of the critical disorder strength
W$_c$ for each supercell  we extrapolate linearly the values of
TDOS at the band center towards zero. Data to be extrapolated are
taken form calculations performed for sets of disorder strengths
in the vicinity of the critical value $W_c \approx 2.1$~\cite{Slevin99,Bulka85,Fehske,song_07,slevin_01,romer_10,romer_11,Mackinnon_81}.

\begin{figure}[h]%[htb]
\includegraphics[width=0.45\textwidth]{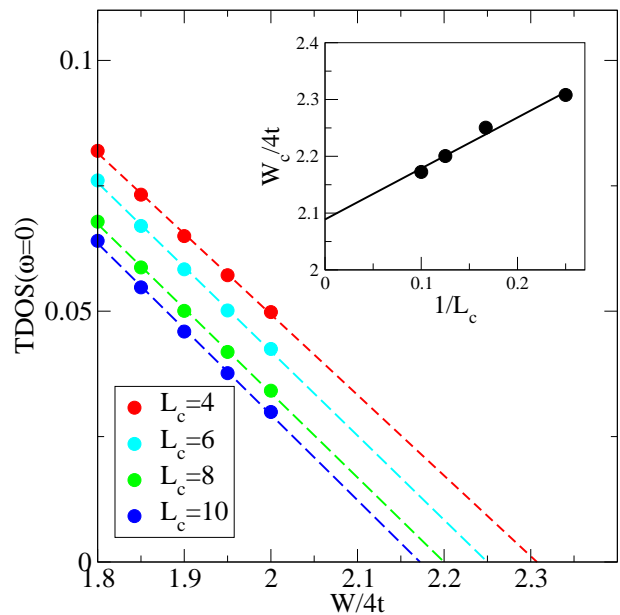}
\caption[]{Extrapolation of TDOS at the band center as a function of disorder strength W for for various supercell sizes with fixed LIZ size $L_{\rm LIZ}=3$. 
Inset: extrapolation of the critical disorder strength $W_c$ to the thermodynamic limit.}
\label{fig:dos_tdos_lc}
\end{figure}

%\subsection{Anderson localization}

In the following we investigate how effective is the typical  embedding scheme in capturing Anderson localization as a consequence of strong disorder. In Fig.~\ref{fig:dos_tdos_lc}, 
%(left panel) we plot the evolution of the average DOS(w) and the typical TDOS(w) density of states at disorder strength $W=2$ for different cluster sizes and fixed $L_{LIZ}=3$.
%We find that the average DOS does not change much as system size increase, at the same time, the TDOS exhibit strong system size dependence and converges slowly as cluster size increases. To study the electron localization in Fig.~\ref{fig:dos_tdos_lc} (right panel) 
we plot the TDOS($\omega=0$) at the band center as function of disorder strength $W$. 
The Anderson transition is then defined by vanishing TDOS($\omega=0$) above the critical disorder strength $W_c$. 
We can do a further extrapolation of W$_c$ vs. $1/L_c$ to 
estimate the critical disorder strength W$_c$ in the 
thermodynamic limit. This is shown in the inset of  Fig.~\ref{fig:dos_tdos_lc}.  The extrapolated value $W_c=2.09$ 
turns out to be in excellent agreement with the exact results 
$W_c \approx 2.10$~\cite{Slevin99,Bulka85,Fehske,song_07,slevin_01,romer_10,romer_11,Mackinnon_81}.

%\section{Conclusion}
In conclusion, we have developed a method for disordered systems, 
which takes the advantage of a local interaction zone (LIZ) 
construction to efficiently compute the local Green's function
corresponding to a supercell embedded into an effective medium. 
We apply this method to a single band 3D Anderson model.
For a typical effective medium embedding of the supercell
we are able to capture the physics of Anderson localization.
The numerical extrapolation predicts an accurate critical 
disorder strength for the localization transition.
We find the embedding method has a quick convergence as the LIZ size, and hence reduces the computational effort.

% By embedding the LIZ in the CPA and TMT effective medium, we demonstrate the important impact of the embedding scheme.
%  We demonstrate that typical medium embedding plays a very important role for capturing the physics of Anderson localization when the embedding scheme is applied. 

The present method may serve  as a guidance for developing an efficient typical medium embedding scheme in the multiple scattering framework. Eventually, this allows first principle
studies of the localization effects in functional materials 
containing disorder. 

%acknowledgements
\textit{Acknowledgments}--
This manuscript is based upon work supported by the U.S. Department of Energy, Office of Science, Office of Basic Energy Sciences under Award Number DE-SC0017861. This work used the high performance computational resources provided by the Louisiana Optical Network Initiative (http://www.loni.org), and HPC@LSU computing.
The work of ME has been supported by U.S. Department of Energy, Office of Science, Basic Energy Sciences, Material Sciences and Engineering Division and it used resources of the Oak Ridge Leadership Computing Facility, which is a DOE Office of Science User Facility supported under Contract DE-AC05-00OR22725.
LC gratefully acknowledge the financial support offered by the Augsburg Center for Innovative Technologies, and by the
Deutsche Forschungsgemeinschaft (DFG, German Research Foundation)
- Projektnummer 107745057 - TRR 80/F6.
\bibliography{LSMS_embedding}
\end{document}